\documentclass{nature}

\usepackage{graphicx}
\usepackage[dvipsnames]{xcolor}
\usepackage{changes}
\usepackage[left=2.0cm, right=2.0cm, top=2.0cm, bottom=2.0cm]{geometry}
\usepackage{tikz}
\usetikzlibrary{shapes}
\usepackage{graphicx}				
\usepackage{dcolumn}		
\usepackage{bm}				
\usepackage{float}
\usepackage{amsmath}
\usepackage{amssymb}
\usepackage{times}
\usepackage{tabularx}
\usepackage{xcolor,colortbl}
\usepackage[11pt]{moresize}
\usepackage{gensymb}
\usepackage{epsfig}
\usepackage{graphicx}
\usepackage{color}
\usepackage{bm}
\usepackage{fixme}
\usepackage{soul}
\usepackage{soulpos}
\usepackage{anysize}
\usepackage{bm}
\usepackage{textcomp}
\usepackage{graphicx}
\usepackage{amsmath,amsfonts}
\usepackage{lineno}
\usepackage{float}
\usepackage{amsfonts}
\usepackage{physics}
\usepackage{upgreek}
\usepackage{textcomp}
\usepackage{xcolor}

\usepackage{xargs}
\newcommandx*{\mypartial}[3] [3]{
 \frac{\partial^{#3} #1} {\partial #2 ^{#3}  }
}

\makeatletter
\let\saved@includegraphics\includegraphics
\AtBeginDocument{\let\includegraphics\saved@includegraphics}

\makeatother
\title{% Elastic and Inelastic Scattering  Among Relativistic Antiferromagnetic Domain Walls\\
Tailoring Elastic Scattering of Relativistic Antiferromagnetic Domain Walls for Collision-based Computing}

\author{Rub\'{e}n M. Otxoa$^{1,2,*}$, G. Tatara$^{3}$, Pierre E. Roy$^{1}$ \& O. Chubykalo-Fesenko$^{4}$}

\begin{document}

\maketitle

\begin{affiliations}
 \item Hitachi Cambridge Laboratory, J. J. Thomson Avenue, CB3 OHE, Cambridge, United Kingdom
 \item Donostia International Physics Center, paseo Manuel de Lardizabal 4, 20018 San Sebasti\'{a}n 20018, Spain
 \item RIKEN  Center  for  Emergent  Matter  Science  (CEMS)  and  RIKEN  Cluster  for Pioneering  Research  (CPR),  2-1  Hirosawa,  Wako,  Saitama,  351-0198  Japan
 \item Instituto de Ciencia de Materiales de Madrid, CSIC, Cantoblanco, 28049 Madrid, Spain
 \end{affiliations}

\begin{abstract}
Soliton-based computing is relied on their unique properties for transporting energy and emerging intact from  head-on collisions. Magnetic domain walls are often referred to as solitons disregarding the strict mathematical definition requiring the above scattering property. Here we demonstrate the conditions of  elastic and inelastic scattering for spin-orbit torque-induced dynamics of antiferromagnetic domain walls  on the example of a technologically relevant Mn$_2$Au material. We show that even  domain walls with opposite winding numbers can experience elastic scattering and we present a corresponding phase diagram as a function of the spin-orbit field strength and duration. The elastic collision requires minimum domain walls speed which we explain assuming an attractive potential created by domain wall pair.  On the contrary, when the domain walls move at lower speeds, their collision is inelastic and results in  a dispersing breather. Our findings will be important for the development soliton-based computing using antiferromagnetic spintronics and we discuss these perspective on our suggestions of how to create NOT and XOR gates. 

\end{abstract}

\maketitle
\section*{Introduction}

As particle-like highly localized  objects, solitons can carry and exchange information, which make them unique entities for unconventional computation\cite{adamatzky2007unconventional,adamatzky2016advances,ziegler2020novel,finocchio2021promise,dieny2020opportunities,grollier2016spintronic,grollier2020neuromorphic}. Robustness to perturbations and very importantly to collisions is an essential ingredient to build soliton-based nanoelectronics. Soliton-based information processing  is typically discussed in relation to their technological applications to nonlinear optics\cite{kivshar2003optical,ablowitz2000optical}, other possibilities such as semiconductor waveguides\cite{aitchison1992observation,trillo2001spatial} or Bose-Einstein condensates also exist\cite{burger1999dark,denschlag2000generating,busch2000motion}. Optical solitons are very localised and fast (typical temporal width corresponding to femto- picosecond timescale) and thus can potentially be used in ultrafast computing and electronics.

Here we suggest another alternative to optical solitons for collision-based computing, based on the use of antiferromagnetic domain walls. Antiferromagnetic  spintronics is currently attracting huge attention due to its energy efficiency, high speed, abundance of materials and invisibility to external magnetic fields\cite{jungwirth2016antiferromagnetic,baltz2018antiferromagnetic,jungwirth2018multiple}. Importantly, antiferromagnetic domain walls have solitonic nature and are not only ultrafast\cite{otxoa2020walker,rama2022inertial} (with velocities up to 40 km/s) but also relativistic and thus contract their length allowing their "charging" by exchange energy and its consequent transport\cite{otxoa2021topologically}. The "discharge" takes place when two domain walls with opposite topological charges collide\cite{otxoa2020walker,otxoa2021topologically}. They can also exhibit multi-cascading processes useful for transporting information from one collision to the next\cite{otxoa2020walker}. 
Being close to solitons and related to the well-known sigma-model \cite{zvezdin1979pis,pitaevskii1980statistical,bar1985dynamics}, antiferromagnetic domain walls of course do not obey the main property of mathematical solitons in exactly integrable systems which state that two (or more) wave packets when collide maintain their shapes. As we show in this article, these conditions, however, can be controlled by strength and duration of current pulses, making them very useful candidates for soliton-based computing.

In nano-magnetism a soliton is interpreted as a spatially localised perturbation whose stability is provided by the magnetisation field structure\cite{bogdanov2001chiral,roessler2006spontaneous,braun2012topological}. Therefore, the elastic collision requirement is relaxed when referring to domain walls (DW), vortices, skyrmions, etc. as solitons. A more accurate way to coin magnetic domain wall is $\textit{kink}$ which connects two ground states but still allows for two objects to annihilate\cite{remoissenet2013waves,rajaraman1982solitons,bishop2012solitons}.
However, for the recombination to occur one needs to attend for the topological character of the two magnetic textures\cite{Manton,kosevich1990magnetic}. Distinct topological textures are characterised by different winding number, $w$, which counts the number of times the magnetisation is wrapped onto itself\cite{braun2012topological}. The winding number density at each point in time $t$, is $w\left(x,t\right)=-\nabla_{x}\phi\left(x,t\right)$ (with the total winding number being $\frac{1}{2\pi}\int w\left(x,t\right)dx$, which for a $\text{180}^{\circ}$ domain wall takes values $\pm\frac{1}{2}$). Here $\phi\left(x,t\right)$ is the in-plane angle of the spin at location $x$  at time $t$ of the spin-configuration along the one dimensional infinite line. Magnetic textures whose winding number (WN) do not change in time are usually called topological solitons\cite{Manton,kosevich1990magnetic}. Topological character of domain walls may be advantageous from the point of view of soliton-based computing since one can naturally assume a kink/antikink to correspond to two information bits.

Strictly speaking a magnetic topological soliton (TMS) requires its winding number to be a constant of motion. However, magnetic textures instabilities such us Walker-Breakdown (WB) for domain walls\cite{mougin2007domain}  or vortex core reversal for vortices\cite{van2006magnetic} lead to a continuous change of the winding number. Moreover, applications of sufficiently strong magnetic fields  annihilate any magnetic texture. Thus, magnetic structures can be referred as topological magnetic solitons only in some limited sense. Here after, we focus  on dynamical regimes far from these  instabilities, i.e. when the topological charge is conserved. In this case, two magnetic textures with opposite winding number can annihilate each other to generate uniform ground state as both spin arrangements belong to the same topological class\cite{braun2012topological} leading to inelastic collision. However, if both domain walls have the same winding number, the recombination is not allowed as there is no continuous deformation to reach the uniform state. Therefore, at first glance, one would expect that elastic collision property would be reserved only for textures with the same winding number elevating their fundamental character to topological solitons. This has been confirmed recently experimentally \cite{thomas2012topological,benitez2015magnetic} and numerically \cite{purnama2014remote,djuhana2009magnetic,dong2015manipulation} for ferromagnetic textures.  The scenario where the collision occurs between two magnetic textures with opposite winding number has always shown a non-soliton character both in experiment\cite{yoshimura2016soliton,togawa2006domain} and simulations so far \cite{kunz2009field,dong2015manipulation} as a DW recombination  was observed. The question we address in this work is whether recombination phenomenon between magnetic textures with opposite winding number can be avoided and if so under what specific conditions. Anticipating our findings, the proper tailoring of the current pulse duration  in antiferromagnetic materials such as Mn$_2$Au can allow the control over elastic and inelastic scattering.  The key property is the velocity achieved by domain walls during the current application. This provides some of the necessary ingredients for soliton-based computing and we discuss the possibilities to create NOT and XOR gates for possible nanoelectronics applications.

\section*{Results and Discussion}
In this work, we demonstrate that antiferromagnetic (AFM) DWs with opposite winding number can preserve their integrity after a head on collision moving at relativistic speeds.
To elucidate the control over the elastic scattering of antiferomagnetic domain walls (AF DWs), we consider a layered antiferromagnet Mn$_2$Au, arranged in a stripe so that DW would have a one-dimensional propagation.  In order to induce magnetisation dynamics in Mn$_2$Au , we make use of the predicted staggered field-like torque in such crystal structures, where the effective magnetic field resulting from a staggered-induced spin-density, $ \text{H}_\text{so}$, possesses opposite signs at each sub-lattice and gives rise to a spin–orbit torque\cite{vzelezny2014relativistic}.  For the description of the atomistic model see Methods section.  The relativistic DW mobility in Mn$_2$Au has been reported elsewhere\cite{otxoa2020walker, rama2022inertial,otxoa2021topologically}. The maximum  AFDW speed, $v_m$ can be obtained from the magnon dispersion relation and is circa 43.3 km/s for Mn$_2$Au\cite{otxoa2020walker}.

In the continuum approximation and taking into account the relative values of anisotropy parameters (see Methods) one can obtain an equation for the in-plane component of the magnetisation Neel vector of the following form
\begin{equation}
\frac{1}{v^2_{\mathrm{m}}} \, \ddot{\varphi}-\left( \partial^2_x \varphi \right)+\frac{1}{2 \Delta^2_0} \, \sin 2 \varphi= - h \, \sin \varphi-\eta \, \dot{\varphi},
\label{eq:1}    
\end{equation}
Here $\Delta_0=\sqrt{a/(8K_{2||})}$ stands for the DW width at rest,   $h=8 \gamma \hbar \, H_{\text{so}}/ \, a$ denotes the reduced scalar spin-orbit (SO) field related to the applied current, $\eta=8\alpha\hbar /a$ describes the DW dissipation, $a=a_0^2(\mathcal{J}_3+|\mathcal{J}_1|/2)$, where $\mathcal{J}_1$ and $\mathcal{J}_3$ are in-plane exchange parameters, $K_{2||}$ is an in-plane anisotropy (see Methods), $a_0$ is the lattice parameter,   $\gamma$ represents the gyromagnetic ratio, $\alpha$ is the Gilbert damping parameter and $\hbar$ is the reduced Planck constant. The l.h.s. of Eq.(\ref{eq:1}) is the famous exactly integrable sine-Gordon equation. Its  soliton (kink)  solution $\varphi=2 \arctan \exp [(x-vt)/\Delta]$ describes the AFDW having velocity $v$ and velocity-dependent width $\Delta=\Delta_0 \sqrt{1-v^2/v_m^2} $. Note that it is also solution of a complete Eq.(\ref{eq:1}) provided that the r.h.s is zero, i.e. AFDW moves with  a stationary velocity  
$v=(\gamma/\alpha) H_{\text{so}}\Delta
\label{Vel}$. 
Thus, AFDW has solitonic nature in the sense that it propagates without changing the form but only  moving at a stationary velocity, i.e. when the energy input provided by the external current is compensated by the dissipation. 

\begin{figure*}
\centering
\includegraphics[scale=0.4]{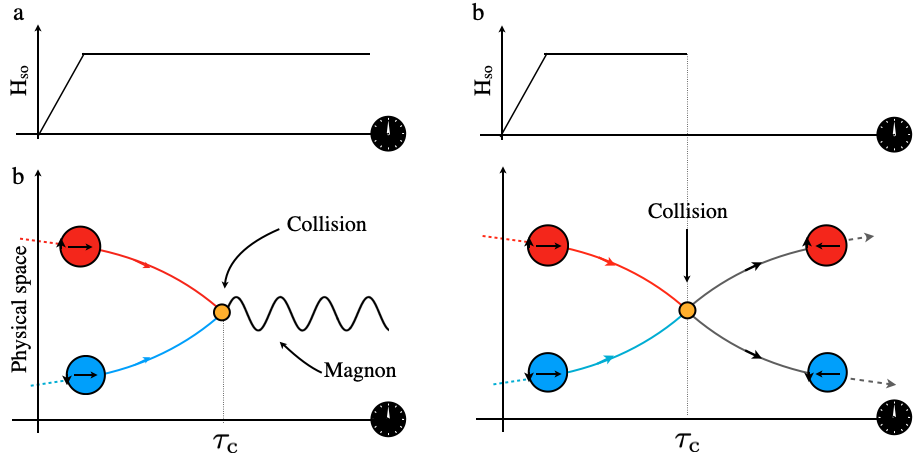}
\caption{\textbf{a} Schematic illustration of two magnetic textures with opposite winding number driven by a SO-field which collide while the SO-field is present. In such a condition the inelastic collision gives rise to the excitation of breather. \textbf{b} Schematic illustration of two magnetic textures with opposite winding number driven by a SO-field which collide elastically in the absence of the SO-field. Each DW preserves its winding number after the collision however, there is a $180^\circ$ phase-shift in the DW's internal spins.}
\label{Fig1}
\end{figure*}

When two solitons of any integrable equation collide, they form a breather. The breather is not a solution of Eq.(\ref{eq:1}) and thus, a priory one cannot expect that AFDWs emerge intact from the collision. In the condition when topological charge is  conserved  and when two AFDWs have the same topological charge, this is indeed the outcome of the collision \cite{}.
When the two topological charges are opposite, we found that two scenarios are possible attending to the duration of the driving mechanism, as  illustrated in Fig.\ref{Fig1}. First, when the SO-field is present for long time, and the collision occurs, the Zeeman energy prevents the DWs to separate as the magnetisation orientation in between the DWs is polarised opposite to the SO-field. The 
 resulting bound state is dispersing in time (Fig.\ref{Fig1}a). In the second case,  one can  switch off the field in the right moment.
Provided that the damping is small (which is indeed the case for magnetic systems) one can expect to find conditions that the resulting breather will not disperse before separating into two kinks (AFDWs).
    
\begin{figure*}
\centering
\includegraphics[scale=0.38]{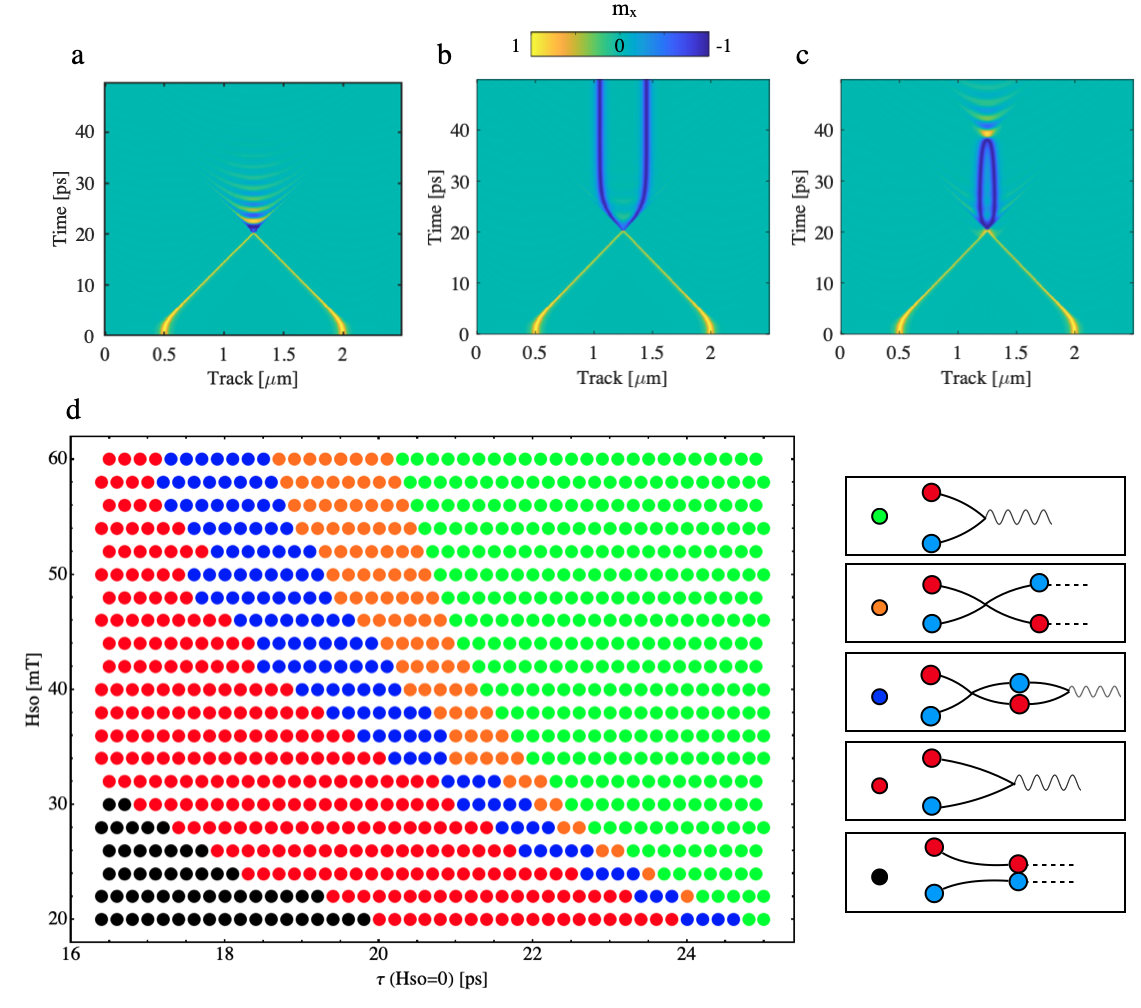}
\caption{\textbf{a} Spatio-temporal evolution of the $s{_\text{x}}$ component of the magnetisation when the SO-field is kept on during the entire simulation. Recombination event occurs and as a result there is a breather. \textbf{b} Spatio-temporal evolution of the $s_{_\text{x}}$ component of the magnetisation when the SO-field is turned off when the collision occurs. The two DWs behave as soliton appearing after the collision with the same winding number but with a $180^\circ$ phase shift in its internal spins. \textbf{c}. Spatio-temporal evolution of the $s_{_\text{x}}$ component of the magnetisation when the SO-field is turned off at 18.5 ps. The two DWs carry $v>v{_\text{e}}$ but the attractive potential after the elastic collision recombine them in the simulation time window. $\textbf{d}$ Phase diagram of the different outcomes from a collision event in terms of the amplitude of the SO-field and its duration.
  }
 \label{Fig2}
\end{figure*}

In what follows we present direct atomistic modelling results of AF DW dynamics under applied current producing SO-field based on Mn$_2$Au complete Hamiltonian (see Methods).
Fig.\ref{Fig2} (a-c) shows the spatio-temporal evolution of magnetic configuration of a system with two DWs  (having opposite winding numbers) colliding under application of  $H{_\text{so}}$= 60 mT that is turned off at different instances. Figure 3 a, shows the case where  an inelastic collision is observed resulting in the breather dispersion. The reason for this is that the $H{_\text{so}}$ has not been switched off after the collision and
 the associated Zeeman energy  prevents AFDWs to separate as the region in between the two DWs have the opposite polarisation to $H{_\text{so}}$.  However, if the SO-field is switched off when the collision occurs (for 60 mT at the time moment around 20.1 ps), then an elastic collision takes place, see Fig. \ref{Fig2}b. We observe a discontinuity in the trajectory of each DW at the moment of the collision and the reappearance of the DW after they tunnel through each other. Importantly, it is observed that while the winding number is preserved through the entire collision process, there has been a  $180^\circ$ phase shift for the DW internal spins for both DWs, i.e. kink has become and antikink.  The third possible scenario Fig.\ref{Fig2}c  results when the two DWs  elastically collide but the distance between them is within the exchange interaction range to observe a recombination in the simulation time-window. The emergence of this breather-like excitation is due to the excess of kinetic energy carried by each DW involved in the collision. Note that, as expected, the resulting breather frequency is monochromatic and independent on the previously applied SO-field.
 Fig. \ref{Fig2}d presents the phase diagram with the possible outcomes observed depending on the magnitude of the SO-field and its duration. The smallest SO-field at which we observe soliton-like collision is 22 mT showing inelastic collision at 20 mT for for all invstigated SO-field duration times.

From now on wards, we refer to AFDW of a given winding number as particle (p) and to that  with opposite winding number as  anti-particle (ap) indistinctly. Hence, one could interpret that each p (ap) behaves as an attractor for its ap (p). For an elastic collision to occur, p(ap) should escape from this potential, i.e. its 
 kinetic energy, K${_\text{p}}$  must be larger than the attractive potential, V${_\text{ap}}$ provided by its own anti-particle. This potential  at a given instance, $t$ is given by the attractive exchange interaction between the two DWs with opposite winding number (see Appendix) and corresponds to
\begin{equation}
E{_\text{p-ap}}=4 A \left( x{_\text{p}} - x{_\text{ap}} \right) \csch \left( \frac{x{_\text{p}}-x{_\text{ap}}}{\Delta} \right),
\end{equation}
where $x_{\text{p}}$ and $x_{\text{ap}}$ represent the particle and anti-particle position with respect to an external and inertial observer.  For sake of simplicity, we will locate the observer at the antiparticle such that $x_{\text{ap}}$=0. Note that this picture assumes that the breather is separated into two "kinks" which is valid when $|x{_\text{p}}-x{_\text{ap}}|\gg 4 \ln{(v/v_m)}$. For relativistic domain walls moving with velocities close to that of the "light"  $c=v_m$, this condition is very loose.

 Hence, we only need to calculate the DW mass \cite{saitoh2004current, feldtkeller1968magnetic, tatara2004theory} in order to obtain the kinetic energy, K${_\text{p}}$
\begin{equation}
K{_\text{p}}=\frac{1}{2} m_{\text{p}} v ^2=\frac{1}{2}\frac{\hbar^2 N}{K_{2\perp}}\frac{v^2}{\Delta^2}
\end{equation}
where  $N=\pi\Delta/a_{0}$ is the number of spins that conform the DW and $K_{2\perp}$ is the second order perpendicular anisotropy, see Methods.   

We consider that the particle has escaped the anti-particle potential when it is at least at a distance of DW width from the anti-particle, i.e. $x{_\text{p}}=$ giving the  escape velocity
\begin{equation}
v_{\text{e}}= \left(\frac{32\pi \,\text{csch}(1) A \delta K_\perp }{ h^2} \right)^{\frac{1}{2}}\Delta\equiv \upxi \Delta(v_e)  .
\end{equation}
where the coefficient $\upxi$ depends on material parameters only but note that  due to the Lorentz invariance the AFDW width on the r.h.s. of this equation is velocity- dependent. Solving explicitly for the escape velocity as a function of SO-field gives: 
\begin{equation}
v_{\text{e}}(H_{\text{so}})=\frac{\upxi \Delta_0 \left[1-\Lambda(H_{\text{so}})\right]^{1/2}}{\left[1+\frac{\upxi^2}{v_m^2}\Lambda(H_{\text{so}})\Delta_0^2\right]^{1/2}}.
\label{eq:escape}
\end{equation}
where
\begin{equation}
\Lambda(H_{\text{so}})=v_m^2\frac{(\gamma H_{\text{so}} \Delta_0)^2 }{(\alpha v_m)^2+ (\gamma H_{\text{so}} \Delta_0)^2}.
\end{equation}
The above equation tells us that the escape velocity depends upon the SO-field and implicitly on the distance the particle needs to {\it{tunnel}} through. 

\begin{figure*}
\centering
\includegraphics[scale=0.4]{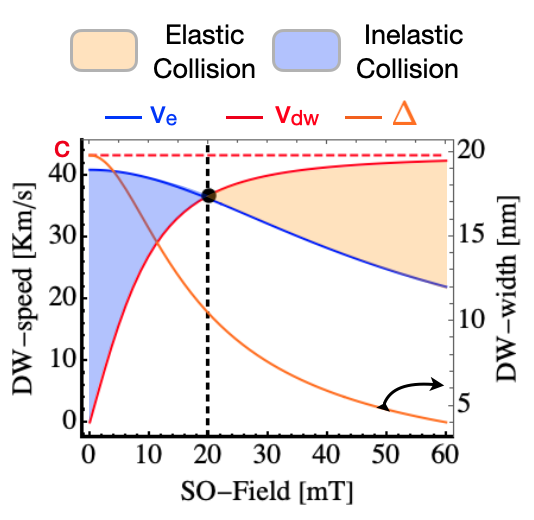}
\caption{DW velocity, $v{_\text{dw}}$, DW width, $\Delta$, and escape velocity, $v{_\text{dw}}$, a as a function of the SO-field. Red dashed line represents the maximum speed velocity extracted from the dispersion relation, $\sim$ 43.3 km/s. For SO-fields larger than the $H{_\text{c}}$ 20 mT, the DW velocity induced by the SO-field is larger than the escape velocity (light orange region). However, for SO-fields lower than $H{_\text{c}}$, the DW can not escape from the attractive potential provided by the other DW as the $v{_\text{dw}} < v{_\text{e}}$ (light blue region).} 
\label{Fig3}
\end{figure*}

In the following, we present our numerical work which corroborates our conclusions. Figure \ref{Fig3} shows the DW speed as a function of the SO-field obtained by our theoretical prediction Eq.(\ref{eq:escape}) (blue line) together with the DW velocity (red line) and its width (orange line) versus SO-field. The last two quantities follow Lorentz increase/contraction.   It can be observed that for SO-fields larger than $H_{cr}\simeq$ 20 mT the actual velocity of the DW is larger than the critical escape velocity and therefore an elastic collision should be observed (Orange filled region). On the contrary for fields smaller than $H_{cr}$, the velocity of the DW is smaller than the escape velocity resulting in both particles being trapped by the attractive potential provided by the other particle. In such a case, an inelastic collision is expected resulting in a DW recombination (purple region). Note that the escape velocity decreases with the increase of the SO field while the DW velocity increases. The intersection seems to be possible for large relativistic velocities only which also favors rapid tunneling and short leaving time of the bound state, providing non-decay conditions. This critical velocity is in a good agreement with what is observed by direct simulations in Fig.\ref{Fig2}. 

 Hence, both our direct modelling and analytical approach demonstrate that the SO-field strength and duration can be tuned the way that  the collision of AFDWs with opposite topological charges would result either in elastic or inelastic outcome. This result can be beneficial for building AF-based future electronic devices for example for constructing possible circuit outputs. In the following we present a possible scenario for two logical operations.  

\section*{Outlook: Proposal for NOT and XOR gates}

Spintronic devices offer a high functionality such as: non-volatile memory, fast operational speeds, well developed routes for writing and reading data, stochastic and even chaotic dynamics for a wide range of magnetic materials \cite{montoya2019magnetization,alvarez2000quasiperiodicity,bertotti2001nonlinear,petit2012commensurability}. Moreover, there has been multiple proposals for logic networks, where the non-volatile nature of the magnetic encoded data would allow for the memory and processing to occur at the same medium. Specifically, topological magnetic solitons (and consequently AFDWs) could be useful for preserving information because of their topological protection. The so-called billiard ball model\cite{durand2002computing,adamatzky2004collision} showed that given a sufficient number of particles, which can collide elastically, any sort of computation can be achieved. Of particular interest is whether such solitons could emulate logic gates. In order to achieve this, a classification is necessary from the evolution space as it cannot be inferred from a local topological rule. As the energy of both logic AFDWs is the same while moving, the logic levels are distinguished by the twofold value of the  associated winding number in the computing region. Interestingly, this system emulates the so-called conservative logic\cite{fredkin1982conservative}, which conserves the physical quantities in which the digital signal is encoded. In particular the WN would be a conserved Boolean quantity. The binary values 1 and 0 are represented by the winding number of a given AFDW. Figure 4 panels a and b show how the winding number is switched after an elastic collision event mimicking a NOT-gate.

However, in order to implement universal logic gates such as XOR-gates one requires to add further complexity to the system. Figure 5 shows a schematic proposal for a XOR-gate which  consists of two free DWs with opposite WNs and a pinned DW with an arbitrary WN. One input to the XOR is associated to magnetic texture's central spin polarization, $m_x>$0 and $m_x<$0 which correspond to logic values 1 and 0 respectively. The second input concerns the SO-field value. The logic value 1 is associated with the critical SO-field, $H>H_\text{crit}$ at which a proliferation of DW-pair with overall WN=0 is observed \cite{.}. The logic value 0 corresponds to $H<H_\text{crit}$, see Figure 5, panels a and b. The output signal after the second collision  (See Figures panels c-f) corresponds to binary values 1 and 0 characterised by the presence or absence of a DW in the output region. Figure 5c shows that after an elastic collision among the two free DWs the resulting DW is accelerated (SO-field increases, see Figure 5a) towards the pinned DW where another elastic collision occurs due to the fact that both DWs have the same WN. The output of this collision is therefore a texture which represents logic value 1. Figure 5d represents once again first an elastic collision among two DWs with opposite WN but, this time after the collision, the SO-field is increased above a value larger than $H_\text{crit}$ resulting in the nucleation of a DW pair with opposite WN. It is known that there is a specific arrangement of the two generated DWs. The one further away from the initial DW has an opposite to it WN \cite{...}. This DW is then accelerated towards the pinned DW but this time, both DWs suffer an inelastic collision and annihilate each other representing a logic value 0. Figure 5e represents the same situation as in Figure 5c but now the pinned layer has opposite polarization than the one in Figure 5c. In this situation, under the action of  the SO-field represented (see Figure 5a), the second collision is inelastic as both DWs have opposite WN resulting in a 0 in the logic output. Figure 5f, shows the same situation as in Figure 5d but with opposite polarization for the pinned layer and same temporal evolution of the SO-field as in panel b. Once again, as the spin-orbit field is lager than $H_\text{crit}$ after the first collision a nucleation event occurs. The generated DW this time has the same WN as the pinned one and the collision is inelastic. Therefore, the output region contains now a magnetic texture. If we associate the presence or absence of a magnetic texture by binary values 1 and 0 respectively,  we obtain the standard XOR-gate logic table, see Figure 5g. We emphasize that the reason why the XOR-gate can be created is due to the conservation of overall WN even when multiple AFDWs are generated. Hence this proposal is related to the so-called conservative logic\cite{fredkin1982conservative} which is chacaterised by the conservation of certain physical quantities. In our case, the overall topological winding number is associated to the spin space.

\section*{Discussion and conclusions}

 In this work we have shown that antiferromagnetic DWs moving at relativistic speed can behave as solitons even in the presence of damping. We have provided a phase diagram for the output two DWs collisions in terms of the SO-field (current) duration and strength. Importantly, the classification these scattering events  cannot be inferred from a local topological rule as we show that the collision of AFDWs with opposite topological charges can produce a variety of different scenarios, including elastic and inelastic outcomes. Therefore, combining soliton intrinsic stability with topological protection provides a very indestructible approach to transmit information or store energy even when collisions occur. So far, studies in the ferromagnets have shown that DWs with opposite winding numbers always recombine. Those studies kept the external magnetic field turned on preventing DWs to separate from each other due to the Zeeman energy penalty. However, a more fundamental reason why DWs in ferromagnets do not show elastic collisions under normal conditions is the fact that the Walker breakdown phenomenon prevents the DWs to move at speeds required to escape the interaction potential. For instance, for parameters extracted from \cite{saitoh2004current}, and typical speed of 100 m/s the corresponding DW kinetic energy is about 5 orders of magnitude lower than the strength of the attractive interaction potential, preventing an elastic collision to occur. Therefore, this soliton character of DWs seems to be reserved only for antiferromagnets irrespective on the overall winding number of the system. Due to the fact that AFDWs can show elastic collision behavior while preserving the overall winding number, we propose and design some basic arrangements where they can act as the building block for an unconventional computing\cite{jakubowski1996can,jakubowski2002computing}.
 
 Comparatively to optical solitons, the AFDWs naturally appear in kink-antikink pair as well as with  two different topological charges. The application of current effectively "charge" them with relativistic exchange energy \cite{otxoa2021topologically} which can be transported on a long distance and be released by inelastic collisions. The elastic collisions can be used for information release. Thus, topological solitons in AF materials are well suited for high speed and dense integrated circuits favorable for technologies based on the application of conservative logic principles of solitons. As shown, logic conservative computation performed by AFDWs does not required any hard-wiring as in conventional computing. Instead, the {\it{wiring}} is provided by the spin medium in the sense that it does not need any fixed hardware structures.
 
Moreover, the efficiency of a computing event is typically given in terms of the quality factor, Q, defined here as the ratio between the energy carried by the AFDWs and the energy dissipated in a computation event. In principle, as the damping is compensated by the action of the SO-field with barely any spin wave emission\cite{otxoa2021topologically} at the moment of the collision event, it is expected the Q-factor to be close to one, as obtained in current CMOS technology. For conventional computers large fraction of the energy associated with certain computation is drawn from the power supply and is dissipated into environment. This  causes a poor waste-heat management\cite{GSP} and acts in detriment in the processing speed. Therefore, the ideal case scenario would correspond to a minimal energy dissipated per computation and the use of spintronics-based AFDW dynamics seems to meet this condition. However, of course,  other technical problems should be solved before this computing scheme could be experimentally realized. In our model, the signal is encoded in the antiferromagnetic DWs and therefore the information travels between logic elements that are fixed spatially.
Certainly, to produce a specific collision between moving AFDWs at a given point, we must have a full control over the distance between the textures, relative topology at the moment of collision and speed at the collision instant etc. Further research that ensures that the corresponding specific conditions are tailored is necessary.

 \newpage
 \section*{Methods}
We consider a magnetic stripe located in xy-plane with a long dimension in x-direction.
 The domain wall dynamics are solved on a  spin-atomic cell structure of Mn$_2$Au.  Mn$_2$Au lattice cell is composed of  totally 10 Mn-sites. In our convection, 4 occupied basal-planes are located in the xy plane with anti-parallel magnetisation orientations along z direction. Upon passing a current along the x-direction, the alternating polarisation of spin-accumulation takes place which gives rise to the corresponding staggered magnetic field, $\textbf{H}_\text{so}$ acting locally on each atomic site and oriented along $\pm$y-direction.   The computational domain is 7500 cells long, one cell wide with periodic boundaries along the stripe width ($y$-direction). The time evolution of a unit vector spin at site $i$, $\textbf{s}_{i}$, is simulated by solving the Landau-Lifshitz-Gilbert equation:
\begin{equation}
\label{eq:LLG}
\frac{d\textbf{s}_{i}}{dt} = -\gamma\,\textbf{s}_{i}\times\textbf{H}_{i}^{\text{eff}}-
\gamma\alpha\,\textbf{s}_{i}\times\left(\textbf{s}_{i}\times\textbf{H}_{i}^{\text{eff}}\right), 
\end{equation} 
where $\alpha$ is the Gilbert damping set here to 0.001 and $\textbf{H}_{i}^{\text{eff}}$ is the effective field resulting from all of the interaction energies. The energies include the three exchange interactions (two antiferromagnetic and one ferromagnetic), magneto crystalline energy contributions and the SO-field. The total energy, $E$ is given by
\begin{eqnarray}
\label{eq:Energy}
E = -2 \sum_{\langle i<j\rangle}{J_{ij}\textbf{s}_{i}\cdot\textbf{s}_{j}}-
K_{2\perp}\sum_{i}{\left(\textbf{s}_{i}\cdot\hat{\textbf{z}}\right)^{2}}- \nonumber
K_{2\parallel}\sum_{i}{\left(\textbf{s}_{i}\cdot\hat{\textbf{y}}\right)^{2}}- \\
\frac{K_{4\perp}}{2}\sum_{i}{\left(\textbf{s}_{i}\cdot\hat{\textbf{z}}\right)^{4}}-
\frac{K_{4\parallel}}{2}\sum_{i}{\left[\left(\textbf{s}_{i}\cdot\hat{\textbf{u}}_{1}\right)^{4}+
\left(\textbf{s}_{i}\cdot\hat{\textbf{u}}_{2}\right)^{4}\right]}-
\mu_{0}\mu_{s}\sum_{i}{\textbf{s}_{i}\cdot\textbf{H}_{i}^{\text{eff}}}.
\end{eqnarray}

The first term on the right-hand side is the exchange energy where $J_{ij}$ is the exchange coefficient along the considered bonds. The second and third terms are the uniaxial hard and easy anisotropies of strengths $K_{2\perp}$ and $K_{2\parallel}$, respectively, while the fourth and fifth terms collectively describes tetragonal anisotropy. For the in-plane part of the tetragonal anisotropy, $\textbf{u}_{1}$=$\left[110\right]$ and $\textbf{u}_{2}$=$\left[1\bar{1}0\right]$. Finally, $\mu_0$ and $\mu_{\text{s}}$ are the magnetic permeability in vacuum and the magnetic moment, respectively. We have used $\mu_{\text{s}}= 4\mu_{B}$ with $\mu_{B}$ being the Bohr magneton (see Table 1 for material parameters). The tetragonal anisotropy was included for sake of completeness as it is present in this material but its role in the high speed dynamics is negligible due to the weak magnitude of its anisotropy constants.
 
\begin{table}
\begin{center}
	\begin{tabular}{|| c c c c c c c||} 
		\hline
		 $J_{i1}k_{\text{B}}^{-1}$[K] & $J_{i2}k_{\text{B}}^{-1}$[K] & $J_{i3}k_{\text{B}}^{-1}$[K] & $K_{2\perp}$[J] & $K_{2\parallel}$[J] & $K_{4\perp}$[J] & $K_{4\parallel}$[J] \\ [0.5ex] 
		\hline\hline
		    -396 & -532 & 115 & -1.303$\times\text{10}^{\text{-22}}$ & 7$K_{4\parallel}$ & 2$K_{4\parallel}$ & 1.855$\times\text{10}^{\text{-25}}$ \\ 
		\hline
	\end{tabular}
\end{center}

\caption{Literature values for material parameters relevant for modelling the spin dynamics. $k_{\text{B}}$ is Boltzmann's constant.}
\label{table:1}
\end{table}

\newpage
\clearpage

\newpage
\clearpage

\newpage
\clearpage

\newpage
\clearpage

\begin{figure*}
\centering
\includegraphics[scale=0.4]{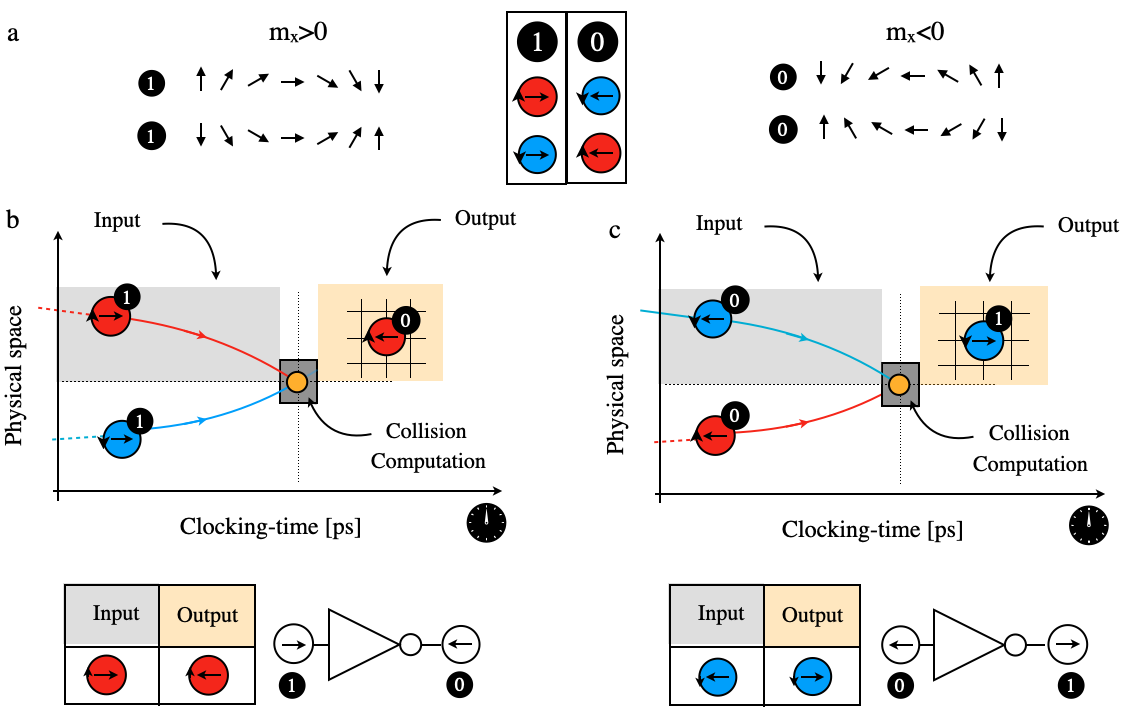}
\caption{Schematic illustration of two magnetic textures with opposite winding number driven by a SO-field which collide elastically in the absence of the SO-field. \textbf{a}. Schematic representation of the logical values associated to different magnetic textures. The value 1 corresponds to a positive $m_x$-component of magnetic texture's central spin whereas the logic value 0 corresponds to a negative $m_x$-component of the magnetic texture's central spin. \textbf{b} Before the collision the AFDW represented as a red circle with a WN$_1$=1 keeps its WN value but the central spin polarization becomes shifted by 180 degrees. Consequently, after the collision event, in the computation region, the elastic collision result is a NOT-gate. \textbf{c} The same occurs when the initial AFDW has an associated WN$_2$=-1 and associated logic value 0 for its central spin polarization. After the collision, while the WN is preserved in the computation region, the phase is shifted 180 degrees, consistent again with NOT-gate operation.}
\end{figure*}

\newpage
\clearpage

\begin{figure*}
\centering
\includegraphics[scale=0.42]{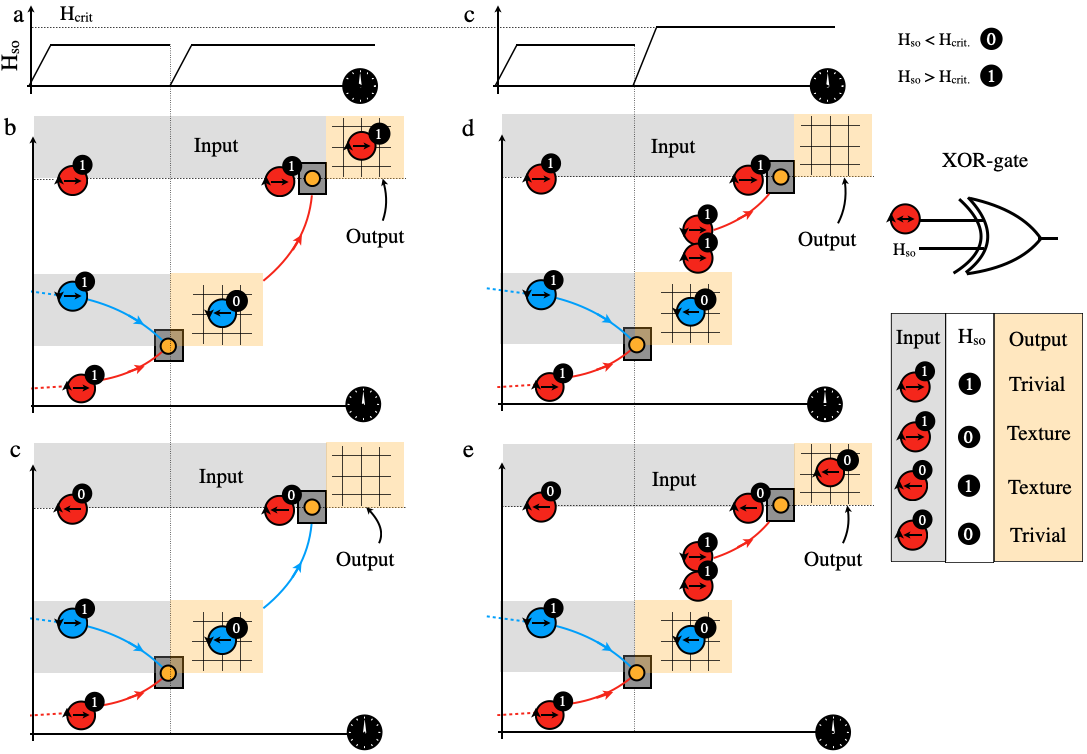}
\caption{Panel \textbf{a} Temporal evolution of the spin-orbit field, H$_\text{so}$. Panel \textbf{b} Temporal evolution of the spin-orbit field, H$_\text{so}$, which will results for a different output when H$_\text{so}>$H$_\text{crit.}$. Panel \textbf{c} shows an elastic collision among two free domain walls (DWs) with opposite winding number,  followed by a second
 elastic collision with a pinned DW in the output region represented in orange. Panel \textbf{d} shows an elastic collision among two free DWs with opposite winding number, followed by the nucleation of another domain wall, with overall WN=0. A second inelastic collision occurs among one of the nucleated DWs and a pinned DW. Panel \textbf{e} Same scenario as in panel \textbf{c} but with a pinned DW with opposite polarization. \textbf{f} same scenario as in panel \textbf{d} but with a pinned DW with opposite polarization.} 
\label{xor}
\end{figure*}

\newpage
\clearpage

\section*{Aknowledgments}
The work of R.M.O. was partially supported by the STSM Grants from the COST Action CA17123 ``Ultrafast opto-magneto-electronics for non-dissipative information technology".

\section*{References}

\bibliography{references}

\end{document}